\documentclass[prb,showpacs,twocolumn]{revtex4}%
\usepackage{amsfonts}
\usepackage{amsmath}
\usepackage{amssymb}
\usepackage{graphicx}
\usepackage{times}%
\providecommand{\U}[1]{\protect\rule{.1in}{.1in}}

\begin{document}
\title{Local spin density in two-dimensional electron gas with hexagonal boundary}
\author{Son-Hsien Chen}
\email{d92222006@ntu.edu.tw}
\affiliation{Department of Physics, National Taiwan University}
\affiliation{National Center for Theoretical Science}
\affiliation{Center for Theoretical Physics, National Taiwan University, Taipei 10617, Taiwan}
\author{Ming-Hao Liu}
\affiliation{Department of Physics, National Taiwan University}
\affiliation{National Center for Theoretical Science}
\affiliation{Center for Theoretical Physics, National Taiwan University, Taipei 10617, Taiwan}
\author{Ching-Ray Chang}
\affiliation{Department of Physics, National Taiwan University}
\affiliation{National Center for Theoretical Science}
\affiliation{Center for Theoretical Physics, National Taiwan University, Taipei 10617, Taiwan}
\date{\today}

\pacs{72.25.Dc, 73.23.--b, 71.70.Ej, 85.75.Nn}

\begin{abstract}
The intrinsic spin-Hall effect in hexagon-shaped samples is investigated. To
take into account the spin-orbit couplings and to fit the hexagon edges, we
derive the triangular version of the tight-binding model for the linear Rashba
[Sov. Phys. Solid State \textbf{2}, 1109 (1960)] and Dresselhaus [Phys. Rev.
\textbf{100}, 580 (1955)] [001] Hamiltonians, which allow direct application
of the Landauer-Keldysh non-equilibrium Green function formalism to
calculating the local spin density within the hexagonal sample. Focusing on
the out-of-plane component of spin, we obtain the geometry-dependent spin-Hall
accumulation patterns, which are sensitive to not only the sample size, the
spin-orbit coupling strength, the bias strength, but also the lead
configurations. Contrary to the rectangular samples, the accumulation pattern
can be very different in our hexagonal samples. Our present work provides a
fundamental description of the geometry effect on the intrinsic spin-Hall
effect, taking the hexagon as the specific case. Moreover, broken spin-Hall
symmetry due to the coexistence of the Rashba and Dresselhaus couplings is
also discussed. Upon exchanging the two coupling strengths, the accumulation
pattern is reversed, confirming the earlier predicted sign change in spin-Hall conductivity.

\end{abstract}
\maketitle

\section{Introduction\label{SecIntro}}

Spintronics, a science investigating the mechanism of electron spins, has
stimulated general theoretical interests in and attempts of industrial
application on how to control electron spins.\cite{App1,App2} By coupling the
electron spin to the charge degrees of freedom, the spin-orbit (SO)
interaction makes manipulating spin electrically possible. In particular, the
spin-Hall effect (SHE), originating from such interaction, has attracted lots
of researchers' attention since it induces a pure transverse spin current
simply as a response to the longitudinal charge current.

According to the sources of SO interaction, two types of the SHE can be
identified: (i) extrinsic SHE\cite{Dyakonov, Hirch, Zhang,Extrinc} with SO
coupling being sourced by the external impurities, and (ii) intrinsic SHE
(ISHE) with inherent SO coupling in the system. In both effects, the SO
interaction subjects the spin-up ($+z$) and spin-down ($-z$) electrons
respectively to opposite forces perpendicular to their transport direction. As
a result, up-spin piles up on one side while down-spin on the other of the
sample (in $x$-$y$ plane), forming the so-called spin-Hall accumulation (SHA).
Inspired by the recent breakthrough in experiments,\cite{Y.K.Kato} these two
effects are intensively under investigation in bulk
semiconductors,\cite{Th1,Th2,Th3,Th4,Th5,Th6,Th7} two-dimensional
semiconductor
heterostructures,\cite{Th8,Th9,Th10,Th11,Sinova1,Sinova2,Nikolic,Nikolic2,Finite1,Finite2,Finite3,SOforce}
as well as the laterally confined quantum wires.\cite{QW1,QW2} Moreover, due
to the stronger magnitude of spin current in the intrinsic case than in the
extrinsic case (several orders greater), the ISHE has attracted intensive
interests in theoretical study since the pioneering
proposals.\cite{Th1,Sinova1}\emph{ }Nevertheless, most investigations focus on
the rectangular geometry. The affection of the boundary in other shapes is
rarely discussed. Characterizing the SHE in finite
system,\cite{Nikolic,Nikolic2,Finite1,Finite2,Finite3,SOforce} this boundary
degree of freedom as well as the varieties in bias configurations offer
alternative ways for manipulating spin, and thus can further generate other applications.

In this paper we study the geometry effect by considering a clean
two-dimensional electron gas (2DEG) with hexagonal shape, which gives us more
degrees of freedom in contacting the sample. Inside the 2DEG, two kinds of
well-known and commonly referred SO couplings are considered: the Rashba SO
(RSO) and the Dresselhaus SO (DSO) couplings. The former is related to the
inversion asymmetry of the structure\cite{Rashba} with its coupling strength
adjustable via the gate voltage;\cite{Nitta,Das2} the latter is caused by bulk
inversion asymmetry\cite{Dresselhaus,Lommer} with coupling strength depending
on the material.\cite{Dyakonov2,Bastard} Employing the Landauer-Keldysh (LK)
Green function method in real space,\cite{Nikolic,Nikolic2,Dattabook} we
analyze the $z$ component SHA pattern inside the hexagon-shaped Landauer setup
with two to six terminals, each of which may be contacted by a semi-infinite
ideal lead. Electrochemical potential differences between these leads induce
longitudinal charge current among terminals. Contrary to the SHE in
rectangular samples with two head-to-tail leads, the SHA patterns in our
hexagonal 2DEG preserve the fundamental spin-Hall symmetry: the spins
accumulate symmetrically (same magnitude) but oppositely (different sign)
about the charge current flow direction, which is in general not straight in
our hexagonal samples. Moreover, presented accumulation patterns are found to
depend on (i) SO interaction strength, (ii) bias strength, (iii) sample size,
(iv) lead configuration, and (v) SO interaction type, among which we put more
emphasis on the last three. For factors (iii) and (v), a SHA reversal effect
(SHARE) is identified and discussed in detail. In particular, upon
interchanging the RSO and DSO coupling strengths for factor (v) the whole SHA
pattern is reversed, confirming the sign change in the spin-Hall conductivity
as previously predicted in Ref. \onlinecite{Sinova2}. As for factor (iv), the
accumulation pattern in certain lead configuration may be very different from
those in rectangular samples.

This paper is organized as follows. In order to apply the real space
tight-binding Hamiltonian in the LK formalism and fit the hexagon edges with
appropriate discrete spatial points, we derive the triangular lattice version
for the linear Rashba and Dresselhaus tight-binding model in Sec.
\ref{TBMsection}. This version also enables further study of the geometry
effect on the SHE with other shapes such as triangle, trapezoid, diamond,
parallelogram, arrow, etc. Section \ref{LKFsection} gives a brief review on
the LK formalism. Numerical results for SHA patterns based on the techniques
introduced in Sec. \ref{TBMsection} and Sec. \ref{LKFsection} in various
hexagonal samples will be reported and discussed in Sec.
\ref{numericalresults}, in the zero temperature limit. We finally conclude in
Sec. \ref{SecConclusion}.

\section{Tight-binding (finite differences) model in triangular lattice with
RSO and DSO couplings\label{TBMsection}}

In this section, we approach the original Hamiltonian by discretizing the
continuous space into lattice-like points with the finite difference method.
The following results are thus reliable as the electron wave length is much
longer than the distance between nearest points. This method exactly treats
the Hamiltonian when the realistic lattice points match the ones in finite
differences, and it is also a very general approach for any lattice
structures. With the constructed triangular lattice, we intend to answer the
essential questions on how the sample boundary, which is basically
controllable in today's technology, affects the spin accumulation.

\begin{figure}
[ptb]
\begin{center}
\includegraphics[
height=2.0228in,
width=2.0176in
]%
{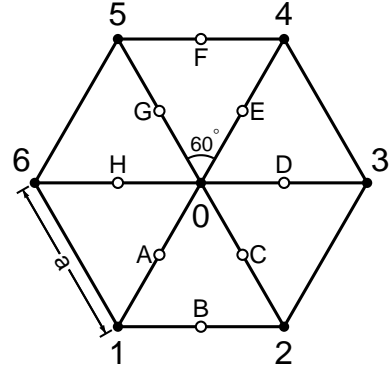}%
\caption{Schematic of the triangular lattice structure. On each lattice point,
the electron has six nearest neighbors to hop.}%
\label{FIG1}%
\end{center}
\end{figure}

Consider a hexagon-shaped 2DEG (set in the $x$-$y$ plane and grown along
[001]) described by the Hamiltonian%
\begin{equation}
\mathcal{H}=\mathcal{H}^{K}+\mathcal{H}^{R}+\mathcal{H}^{D}%
,\label{Hamiltonian}%
\end{equation}
where $\mathcal{H}^{K}=p^{2}/2m^{\star}$ (with $m^{\star}$ being the electron
effective mass) is the kinetic energy, and%
\begin{align}
\mathcal{H}^{R} &  =\frac{\alpha}{\hbar}\left(  p_{y}\sigma_{x}-p_{x}%
\sigma_{y}\right)  \label{HR}\\
\mathcal{H}^{D} &  =\frac{\beta}{\hbar}\left(  p_{x}\sigma_{x}-p_{y}\sigma
_{y}\right)  \label{HD}%
\end{align}
denote the RSO and DSO Hamiltonians with coupling strengths $\alpha$ and
$\beta$, respectively. Here $p_{x}$ and $p_{y}$ are the two independent
contributions to the two-dimensional momentum operator: $\vec{p}=\left(
p_{x},p_{y}\right)  $, and the Pauli matrices $\vec{\sigma}=\left(  \sigma
_{x},\sigma_{y},\sigma_{z}\right)  $ are used. To describe the system in real
space, one can apply the tight-binding method to extract $\mathcal{H}$ in the
real space lattice basis. However, to fit the hexagonal boundary, one has to
discretize the space in triangular lattice structure, as shown in Fig.
\ref{FIG1}, where the lattice points are labeled as $0,1,\cdots,6$, while
$A,B,\cdots,G$ are auxiliary points.
Using the finite difference method\cite{Dattabook} and labeling the wave
function at position $\mathbf{r}$ as $\psi_{\mathbf{r}},$ the general
expressions for first and second derivatives are approximated by%
\begin{align}
\left.  \frac{\delta\mathbf{r}}{\delta r}\cdot\frac{d}{d\mathbf{r}}%
\psi_{_{\mathbf{r}}}\right\vert _{\mathbf{r}=\mathbf{\bar{r}}} &
=\frac{\left(  \psi_{\mathbf{r}_{1}}-\psi_{\mathbf{r}_{2}}\right)  }{\delta
r},\label{1std}\\
\left.  \left(  \frac{\delta\mathbf{r}}{\delta r}\cdot\frac{d}{d\mathbf{r}%
}\right)  ^{2}\psi_{_{\mathbf{r}}}\right\vert _{\mathbf{r}=\mathbf{\bar{r}}}
&  =\frac{\left(  \psi_{\mathbf{r}_{1}}+\psi_{\mathbf{r}_{2}}-2\psi
_{\mathbf{\bar{r}}}\right)  }{\left(  \delta r/2\right)  ^{2}},\label{2std}%
\end{align}
with $\mathbf{\bar{r}\equiv}\left(  \mathbf{r}_{1}+\mathbf{r}_{2}\right)  /2$
and $\delta\mathbf{r\equiv r}_{1}-\mathbf{r}_{2}$. Thus the operation of these
momentum operators can be expressed as%
\begin{subequations}
\begin{align}
p_{x}\psi_{0} &  =\frac{-i\hbar}{a}\left(  \psi_{H}-\psi_{D}\right)
\label{px}\\
p_{y}\psi_{0} &  =\frac{-i\hbar}{\sqrt{3}a}\left(  \psi_{F}-\psi_{B}\right)
,\label{py}%
\end{align}
with $a$ being the lattice constant. Note the different denominators in Eqs.
(\ref{px}) and (\ref{py}) due to the different distances between
$\overline{HD}$ and $\overline{FB}$. In order to construct the matrix
representation of the Hamiltonian in the real space triangular lattice basis,
our goal is to express the operation of $\mathcal{H}$ on the wave function at
point $0$: $\psi_{0}$, in terms of those on its nearest neighbors: $\psi
_{1},\psi_{2},\cdots,\psi_{6}$.

Starting with the kinetic energy term, we take advantage of the isotropy of
$\mathcal{H}^{K},$ which implies that the six directions (from site 0 to
$1,2,\cdots,6$) in $\mathcal{H}^{K}$ should take the same form. From Eq.
(\ref{1std}), we therefore have two additional momentum operators $p_{1}$
along $\overline{EA}$ and $p_{2}$ along $\overline{GC}$ with their operations
on $\psi_{0}$ given by%
\end{subequations}
\begin{subequations}
\begin{align}
p_{1}\psi_{0}  &  =\frac{-i\hbar}{a}\left(  \psi_{E}-\psi_{A}\right)
\label{p1}\\
p_{2}\psi_{0}  &  =\frac{-i\hbar}{a}\left(  \psi_{G}-\psi_{C}\right)  .
\label{p2}%
\end{align}
Combining the three equal contributions from $p_{x},$ $p_{1},$ and $p_{2}$ [by
using Eqs. (\ref{px}), (\ref{p1}), and (\ref{p2})] and defining the hopping
energy $t_{0}\equiv\hbar^{2}/2m^{\star}a^{2}$, straightforward substitution
and rearrangement give%
\end{subequations}
\begin{align}
&  \frac{p_{1}^{2}+p_{2}^{2}+p_{x}^{2}}{2m^{\star}}\psi_{0}\nonumber\\
&  =-t_{0}\left[  \left(  \psi_{1}+\psi_{4}-2\psi_{0}\right)  +\left(
\psi_{2}+\psi_{5}-2\psi_{0}\right)  \right. \nonumber\\
&  \left.  +\left(  \psi_{3}+\psi_{6}-2\psi_{0}\right)  \right]
\label{p1^2+p2^2+px^2}\\
&  =-t_{0}\left[  \left(  \psi_{2}+\psi_{4}-2\psi_{D}\right)  +\left(
\psi_{1}+\psi_{5}-2\psi_{H}\right)  \right. \nonumber\\
&  \left.  +\left(  \psi_{3}+\psi_{6}-2\psi_{0}\right)  \right]
-2t_{0}\left(  \psi_{D}+\psi_{H}-2\psi_{0}\right)  . \nonumber\label{p1p2px}%
\end{align}
From the above equation, together with Eq. (\ref{2std}), we deduce $\left[
\left(  p_{1}^{2}+p_{2}^{2}+p_{x}^{2}\right)  /2m^{\star}\right]  \psi
_{0}=\left(  3/2\right)  \left[  \left(  p_{x}^{2}+p_{y}^{2}\right)
/2m^{\star}\right]  \psi_{0},$ and finally obtain%
\begin{equation}
\mathcal{H}^{K}=\frac{2}{3}\frac{\left(  p_{1}^{2}+p_{2}^{2}+p_{x}^{2}\right)
}{2m^{\star}}. \label{HKtri}%
\end{equation}
Note that here the long wave length limit $p_{y}^{2}\left(  \psi_{D}+\psi
_{H}\right)  /2\approx p_{y}^{2}\left(  \psi_{0}\right)  $ is assumed.%
\begin{ruledtabular}\begin{table}[<t>]%
\begin{tabular}
[c]{cccc}%
index $\left(  i,j\right)  $ & $\mathcal{H}_{\left(  i,j\right)  }^{K}$ &
$\mathcal{H}_{\left(  i,j\right)  }^{R}$ & $\mathcal{H}_{\left(  i,j\right)
}^{D}$\\\hline
(0,0) & $4t_{0}$ & $0$ & $0$\\
(0,1) & $-\dfrac{2}{3}t_{0}$ & $\dfrac{it^{R}}{\sqrt{3}}\sigma_{x}%
-\dfrac{it^{R}}{2}\sigma_{y}$ & $-\dfrac{it^{D}}{\sqrt{3}}\sigma_{y}%
+\dfrac{it^{D}}{2}\sigma_{x}$\\
(0,2) & $-\dfrac{2}{3}t_{0}$ & $\dfrac{it^{R}}{\sqrt{3}}\sigma_{x}%
+\dfrac{it^{R}}{2}\sigma_{y}$ & $-\dfrac{it^{D}}{\sqrt{3}}\sigma_{y}%
-\dfrac{it^{D}}{2}\sigma_{x}$\\
(0,3) & $-\dfrac{2}{3}t_{0}$ & $\dfrac{it^{R}}{2}\sigma_{y}$ & $-\dfrac
{it^{D}}{2}\sigma_{x}$\\
(0,4) & $-\dfrac{2}{3}t_{0}$ & $-\dfrac{it^{R}}{\sqrt{3}}\sigma_{x}%
+\dfrac{it^{R}}{2}\sigma_{y}$ & $\dfrac{it^{D}}{\sqrt{3}}\sigma_{y}%
-\dfrac{it^{D}}{2}\sigma_{x}$\\
(0,5) & $-\dfrac{2}{3}t_{0}$ & $-\dfrac{it^{R}}{\sqrt{3}}\sigma_{x}%
-\dfrac{it^{R}}{2}\sigma_{y}$ & $\dfrac{it^{D}}{\sqrt{3}}\sigma_{y}%
+\dfrac{it^{D}}{2}\sigma_{x}$\\
(0,6) & $-\dfrac{2}{3}t_{0}$ & $-\dfrac{it^{R}}{2}\sigma_{y}$ & $\dfrac
{it^{D}}{2}\sigma_{x}$%
\end{tabular}
\caption{Matrix elements of the kinetic energy term $\mathcal{H}%
^{K}$, the linear
Rashba term $\mathcal{H}^{R}$, and the linear Dresselhaus term $\mathcal
{H}^{D}$ in
Eq. (\ref{Hamiltonian}) in the triangular version of tight-binding model.}
\label{Hij}\end{table}\end{ruledtabular}%

Next we seek for the matrix representation for $\mathcal{H}^{R}$ and
$\mathcal{H}^{D}$. Similar to the expression for $\mathcal{H}^{K}$, we need to
express the operations $p_{x}\psi_{0}$ and $p_{y}\psi_{0}$ in terms of
$\psi_{1},$ $\psi_{2},\cdots,\psi_{6}$. This can be done by noting
\begin{align}
p_{x}\frac{\psi_{B}+\psi_{F}}{2}  &  =\frac{-i\hbar}{a}\left[  \left(
\psi_{2}-\psi_{1}\right)  +\left(  \psi_{4}-\psi_{5}\right)  \right]  \approx
p_{x}\psi_{0}\label{px2}\\
p_{y}\frac{\psi_{D}+\psi_{H}}{2}  &  =\frac{-i\hbar}{\sqrt{3}a}\left[  \left(
\psi_{4}-\psi_{2}\right)  +\left(  \psi_{5}-\psi_{1}\right)  \right]  \approx
p_{y}\psi_{0}, \label{py2}%
\end{align}
where the long wave length limit is again assumed. Thus the operation on
$\psi_{0}$ due to terms required in the linear Rashba and Dresselhaus
Hamiltonians given in Eqs. (\ref{HR}) and (\ref{HD}) can be completely
expressed in terms of $\psi_{1},$ $\psi_{2},\cdots,\psi_{6}$.

Using Eq. (\ref{HKtri}) with Eq. (\ref{p1^2+p2^2+px^2}) and Eqs.
(\ref{px2})--(\ref{py2}), we obtain the finite difference representation of
the Hamiltonian $\left(  \mathcal{H}\psi\right)  _{0}=\sum_{n=1}%
^{6}\mathcal{H}_{\left(  0,n\right)  }\psi_{n}$. Since the matrix elements are
non-vanishing only for nearest hopping:%
\begin{align*}
\mathcal{H}_{\left(  i,j\right)  }  &  =\mathcal{H}_{\left(  0,0\right)
},\text{ if }i=j\\
&  =\mathcal{H}_{\left(  0,n\right)  },\text{ }n\in\{1,2,..6\},\text{ if
}i,j\text{ are nearest neighbors}\\
&  =0,\text{ otherwise,}%
\end{align*}
this expression is equivalent to the tight-binding model $\mathcal{H}%
^{\text{TBM}}=\sum_{i,j}\mathcal{H}_{\left(  i,j\right)  }c_{i}^{\dagger}%
c_{j}$, with $c^{\dagger}$ ($c$) being the creation (annihilation) operator.
Defining the RSO and DSO hopping parameters as $t^{R}\equiv\alpha/2a$ and
$t^{D}\equiv\beta/2a$, respectively, the on-site energy $\mathcal{H}_{\left(
0,0\right)  }$ and hopping matrix elements $\mathcal{H}_{\left(  0,n\right)
}$ are summarized in Table \ref{Hij}, with each row corresponding to different
hopping direction referring to Fig. \ref{FIG1}. For example, the first row,
labeled by $\left(  0,0\right)  $ represents the on-site energy or the energy
$\mathcal{H}_{\left(  0,0\right)  }$ hopping from site 0 to 0, while the
second row, labeled by $\left(  0,1\right)  $, represents the energy
$\mathcal{H}_{\left(  0,1\right)  }$ hopping from site 1 to 0 so that Table
\ref{Hij} gives $\mathcal{H}_{\left(  0,0\right)  }=4t_{0}$ and $\mathcal{H}%
_{\left(  0,1\right)  }=\mathcal{H}_{\left(  0,1\right)  }^{K}+\mathcal{H}%
_{\left(  0,1\right)  }^{R}+\mathcal{H}_{\left(  0,1\right)  }^{D}%
=-2t_{0}/3+(it^{R}\sigma_{x}/\sqrt{3}-it^{R}\sigma_{y}/2)+(it^{D}\sigma
_{y}/\sqrt{3}-it^{D}\sigma_{x}/2)$, etc$\emph{.}$

\section{Landauer-Keldysh Formalism\label{LKFsection}}

In this section we give a brief review on the LK formalism, namely, the
Keldysh non-equilibrium Green function
formalism\cite{Keldysh,Caroli,KeldyshDerivation,IntroToKeldysh} applied on
Landauer multiterminal setups. The following review, which is in general
applicable for mesoscopic systems with any shapes, is mainly based on Ref. \onlinecite{Nikolic2}.

Consider a hexagon-shaped Landauer setup, each side of which may be contacted
by a semi-infinite two-dimensional ideal lead. In the following calculation,
the leads are assumed to be in their own thermal equilibrium all the time
(before and after contact), while the hexagonal conductor is in a
non-equilibrium state, meaning that the electrons therein are not distributed
according to a single Fermi-Dirac distribution. The occupation number of
electron with spin $\sigma$ on site $\mathbf{m}$ at time $t$ is determined by
the diagonal matrix element $G_{\mathbf{mm,\sigma\sigma}}^{<}(t,t;t_{s})$ of
the lesser Green function%
\begin{equation}
G_{\mathbf{m}^{\prime}\mathbf{m,\sigma}^{\prime}\sigma}^{<}\left(  t^{\prime
},t;t_{s}\right)  =\frac{i}{\hbar}\left\langle c_{\mathbf{m\sigma}}^{\dag
}\left(  t,t_{s}\right)  c_{\mathbf{m}^{\prime}\sigma^{\prime}}\left(
t^{\prime},t_{s}\right)  \right\rangle , \label{Lesser G}%
\end{equation}
where $c_{\mathbf{m}\sigma}^{\dag}$($c_{\mathbf{m}\sigma}$) is the creation
(annihilation) operator for the corresponding site $\mathbf{m}$ and spin index
$\sigma$. The lesser Green function given in Eq. (\ref{Lesser G}) in general
depends on the switching time $t_{s}$, at which the leads are brought into
contact. (In the following expressions, $t_{s}$ will be set to $-\infty$.) In
the steady state, however, only the relative time variable $t^{\prime}%
-t\equiv\tau$ is relevant, so that the lesser Green function can be expressed
in terms of its Fourier components $G_{\mathbf{m}^{\prime}\mathbf{m,\sigma
}^{\prime}\sigma}^{<}\left(  \tau\right)  =\left(  2\pi\hbar\right)  ^{-1}%
\int_{-\infty}^{\infty}dEG_{\mathbf{m}^{\prime}\mathbf{m,\sigma}^{\prime
}\sigma}^{<}\left(  E\right)  e^{iE\tau/\hbar}$, leading to the occupation
number%
\begin{equation}
\left\langle c_{\mathbf{m}}^{\dag}c_{\mathbf{m}}\right\rangle =\frac{\hbar}%
{i}G_{\mathbf{m},\mathbf{m}}^{<}\left(  \tau=0\right)  =\frac{1}{2\pi i}%
\int_{-\infty}^{\infty}dEG_{\mathbf{mm,\sigma\sigma}}^{<}\left(  E\right)  .
\end{equation}

The solution to the lesser Green function matrix $\mathbf{G}^{<}\left(
E\right)  $ is a kinetic equation $\mathbf{G}^{<}\left(  E\right)
=\mathbf{G}^{R}\left(  E\right)  \mathbf{\Sigma}^{<}\left(  E\right)
\mathbf{G}^{A}\left(  E\right)  ,$with $\mathbf{G}^{R}\left(  E\right)
=\left[  E\mathbf{I}-\mathbf{H}_{\text{eff}}\right]  ^{-1}$ and $\mathbf{G}%
^{A}\left(  E\right)  =\left[  \mathbf{G}^{R}\left(  E\right)  \right]
^{\dagger}$ being the retarded and advanced Green functions, respectively. The
interactions between the conductor and the leads are included in the
self-energy
\begin{equation}
\left[  \mathbf{\Sigma}_{p}\right]  _{\left(  i,j\right)  }=\left[
t^{2}\mathbf{g}_{p}^{R}\left(  E\right)  \right]  _{\left(  p_{i}%
,p_{j}\right)  },\label{Sigma}%
\end{equation}
which yields the effective conductor Hamiltonian $\mathbf{H}_{\text{eff}%
}=\mathbf{H}+\sum\nolimits_{p}\mathbf{\Sigma}_{p}(E-eV_{p})$, where
$\mathbf{H}$ is the conductor Hamiltonian with matrix elements $\mathcal{H}%
_{\left(  i,j\right)  }$ summarized in Table \ref{Hij}. In Eq. (\ref{Sigma}),
the indices $\left(  i,j\right)  $ on the left side and $\left(  p_{i,}%
p_{j}\right)  $ on the right side label the adjacent sites connecting the
conductor and the lead $p$, and $\mathbf{g}_{p}^{R}\left(  E\right)  $ is the
retarded Green function for the corresponding isolated lead. The lesser
self-energy $\mathbf{\Sigma}^{<}$, which generates the open channels, is given
by
\begin{align}
\mathbf{\Sigma}^{<}\left(  E\right)   &  =-%
{\displaystyle\sum\limits_{p}}
\left[  \mathbf{\Sigma}_{p}(E-eV_{p})-\mathbf{\Sigma}_{p}^{\dagger}%
(E-eV_{p})\right]  \nonumber\\
&  \times f\left(  E-eV_{p}\right)  ,\label{Smls}%
\end{align}
where $f\left(  E-eV_{p}\right)  $ is the Fermi-Dirac distribution and $V_{p}$
is the applied bias voltage on lead $p$. Note that Eq. (\ref{Smls}) implies
that $\mathbf{G}^{<}$ is an anti-Hermitian matrix ensuring that the occupation
number is a real quantity, while the non-Hermitian Hamiltonian $\mathbf{H}%
_{\text{eff}}$ yields the finite eigenstate lifetime of quasi-particle.

\section{Numerical results\label{numericalresults}}

In this section we present numerical results for the $z$ component of the
local spin density $\left\langle S_{z}\right\rangle $ inside the hexagonal
sample, made of InGaAs/InAlAs heterostuctures grown along [001]. Typical
parameters: the electron effective mass $m^{\star}=0.05m_{e}$ ($m_{e}$ is the
electron rest mass) and the lattice constant $a=3$ $%
\operatorname{nm}%
$ (yielding the hopping energy $t_{0}=84.68$ m$%
\operatorname{eV}%
$), will be taken,\cite{Nitta} and the Fermi energy $E_{F}=-3.8t_{0}$ close to
the band bottom $E_{b}=-4t_{0}$ will be chosen (so that the long wave length
approximation is valid). The size of the regular hexagonal samples will range
from $N=8$ to $N=20$, $N$ being the number of sites per hexagon edge, while
the applied bias on the leads will be set either $eV_{p}=+eV/2$,
$eV_{p}=-eV/2$, or $eV_{p}=0$, to be denoted shortly as \textquotedblleft%
$+$\textquotedblright, \textquotedblleft$-$\textquotedblright, and
\textquotedblleft0\textquotedblright\ in each presented figure, respectively.
Note that here the electron is negatively charged as $e=-\left\vert
e\right\vert $, so that $eV_{p}>0$ implies $V_{p}<0$, meaning that electrons
will flow from \textquotedblleft$+$\textquotedblright\ to \textquotedblleft%
$-$\textquotedblright, in all the presented figures. We will take
$eV=0.4t_{0}$ and $eV=10^{-3}t_{0}$, which will be referred to as high bias
and low bias, respectively. Discussion with Rashba-type 2DEGs, for which
$t^{R}=0.1t_{0}$ and $t^{D}=0$ are set, will be first addressed. Not until
Sec. \ref{SymmetryBreaking:RDcoexist} will we change $t^{R}$ and turn on
$t^{D}$. Note that in most of our numerical results, we put parameters
identical with Ref. \onlinecite{Nikolic}, so that direct comparison and
correspondence can be clearly identified.

\subsection{SHA: General feature\label{SHA:general}}

\begin{figure}
[ptb]
\begin{center}
\includegraphics[
height=3.3615in,
width=3.1765in
]%
{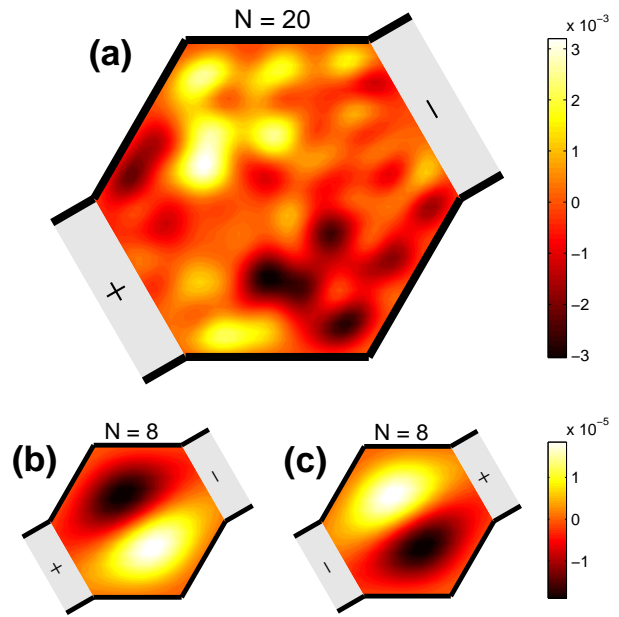}%
\caption{(Color online) Non-equilibrium spin-Hall accumulation $\left\langle
S_{z}\right\rangle $ in a (a) $N=20$ sample under high bias, and a (b) $N=8$
sample under low bias. (c) is the same as (b), except the reversed bias. The
color bars calibrate $\left\langle S_{z}\right\rangle $ in units of $\hbar/2$,
and $\pm$ signs indicate the applied bias $\pm eV/2$ on the corresponding
lead.}%
\label{FIG2}%
\end{center}
\end{figure}

As mentioned in Sec. \ref{SecIntro}, the SHA pattern in hexagon-shaped samples
can be very different from those in rectangular ones. Beginning with Fig.
\ref{FIG2}(a), we plot $\left\langle S_{z}\right\rangle $ in a $N=20$ sample
with two head-to-tail leads under high bias. The accumulation pattern therein
clearly shows the geometry effect: the edge peaks of the SHA follow the shape
of the sample. Moreover, due to the head-to-tail lead configuration, the
accumulation pattern is perfectly spin-Hall-symmetric about the central axis
connecting the two leads: spin accumulation on one side is the mirror (with
different sign) of that on the other.%

When reducing the sample size to $N=8$, one can see that the accumulation
pattern changes sign, as shown in Fig. \ref{FIG2}(b), where low bias is
applied. This is because the spin-orbit force\cite{SOforce} depends on not
only the sign of $S_{z}$, but also the system size relative to the Rashba spin
precession length\cite{HaoPL}%
\begin{equation}
L_{\text{SO}}\equiv\frac{\pi\hbar^{2}}{2m^{\star}\alpha}=\frac{\pi at_{0}%
}{2t^{R}}, \label{LSO}%
\end{equation}
which is the distance required by the RSO interaction to rotate the spin by
$\pi$. With $t^{R}=0.1t_{0}$ here, Eq. (\ref{LSO}) yields the precession
length $L_{\text{SO}}=5\pi a\approx15.7a$. We will address this accumulation
reversal effect a bit further later. Upon reversing the bias, Fig.
\ref{FIG2}(c) shows a flipped pattern, as expected. In short, Fig. \ref{FIG2}
can be regarded as the hexagonal version of Fig. 1 shown in Ref.
\onlinecite{Nikolic},\ and the general feature of the intrinsic SHE is preserved.

\subsection{SHA: Symmetry investigation\label{SHA:further}}

From now on we focus on the low-bias regime and, in this subsection, we put
emphasis on different lead configurations. Various $N=12$ samples with 2
leads, 4 leads, and 6 leads with different configurations will be shown.%
\begin{figure}
[ptb]
\begin{center}
\includegraphics[
trim=1.727336in 0.000000in 0.000000in 0.000000in,
height=4.9761in,
width=2.7475in
]%
{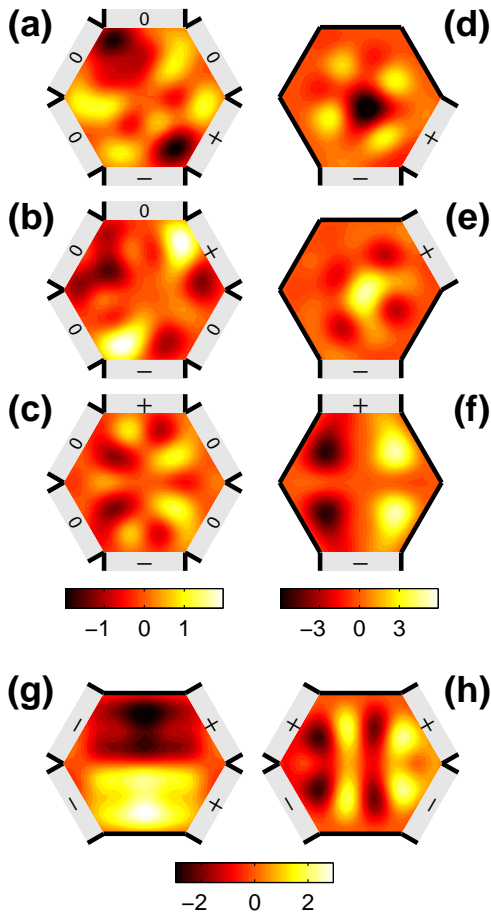}%
\caption{(Color online) Spin-Hall accumulation in $N=12$ hexagonal samples
with (a)--(c) 6 leads, (d)--(f) 2 leads, and (g)--(h) 4 leads, under low bias.
In addition to the $\pm$ signs indicating the applied $\pm eV/2$ bias, the
zero signs in (a)--(c) label the unbiased leads. Color bars calibrate
$\left\langle S_{z}\right\rangle $ in units of $\left(  \hbar/2\right)
\times10^{-5}$.}%
\label{FIG3}%
\end{center}
\end{figure}

Beginning with the 6-lead cases, Figs. \ref{FIG3}(a)--\ref{FIG3}(c) show SHA
patterns under a series of different biasing. For the non-head-to-tail biasing
in Figs. \ref{FIG3}(a) and \ref{FIG3}(b), one can see a curved spin-Hall
symmetry which roughly follows the charge current flow. Moreover, such a bent
SHA may induce $\left\langle S_{z}\right\rangle $ on edges with not
necessarily the opposite signs, e.g., see the dark spots in the left-top and
right-bottom corners shown in Fig. \ref{FIG3}(a). For the head-to-tail-biased
sample of Fig. \ref{FIG3}(c), the straight spin-Hall symmetry is recovered, as
expected. As a direct correspondence, we unplug those unbiased electrodes and
show the accumulation patterns in Figs. \ref{FIG3}(d)--(f). Similar properties
mentioned above are still valid, but each pattern becomes totally different,
implying that the spin accumulation is sensitively affected by the
configuration of attached leads.

We now focus on Figs. \ref{FIG3}(c) and \ref{FIG3}(f) and discuss the SHARE
for these two distinct cases: open vs. closed boundaries. For the open case
Fig. \ref{FIG3}(c), one can clearly observe that the accumulation pattern
reverses with different transverse length, which ranges from $\left(
N-1\right)  a=11a$ (the hexagon edge length) to $2\left(  N-1\right)  a=22a$
(the maximal length of the hexagon) here and covers the precession length
$L_{\text{SO}}=15.7a$. Near the biased top and bottom electrodes, the spin
accumulation signs can be well explained by the semiclassical SO force (to be
discussed further later) proportional to $\left(  \mathbf{p}\times
\mathbf{e}_{z}\right)  \otimes\sigma_{z}$ [see Eq. (\ref{FSO}) in Sec.
\ref{SymmetryBreaking:RDcoexist}], which predicts right deflection for
$\sigma_{z}=+1$ and left deflection for $\sigma_{z}=-1$. At central regions
where the transverse length exceeds $L_{\text{SO}}$, the Rashba spin
precession turns $\sigma_{z}=\pm1$ into $\sigma_{z}=\mp1$ and thus flips the pattern.

For the closed case Fig. \ref{FIG3}(f), the SHARE is slightly different, as
one can see that the bright and dark regions do not swap locally. A basic
difference between the open and closed cases is that the electron does not
have the outflow degree of freedom in the latter case. Therefore, the
criterion of whether the pattern is flipped or not lies on the maximum width
of the whole sample, which amounts to $22a$ for this $N=12$ case and exceeds
$L_{\text{SO}}$ already. Accordingly, the SO force predicts right (left)
deflection for $\sigma_{z}=-1$ ($\sigma_{z}=+1$) electrons, in agreement with
Fig. \ref{FIG3}(f). Indeed, the critical size for the closed sample (with two
head-to-tail leads) to exhibit the flipped accumulation patterns can be shown
as $N=9$, but we do not further show here.

Figures \ref{FIG3}(g) and \ref{FIG3}(h) show accumulation patterns in the
hexagon sample attached by 4 leads with two different bias configurations. As
in the former case longitudinal bias is arranged similar to the rectangular
sample attached to two leads, regular spin-Hall accumulation is observed in
Fig. \ref{FIG3}(g). Interestingly, when we rearrange the applied bias as that
in Fig. \ref{FIG3}(h), \textquotedblleft local\textquotedblright\ SHA\ is
obtained: spin accumulation in the whole sample is divided into two parts,
subject to the two separate pairs of positive-negative electrodes. The reason
to obtain such a divided accumulation pattern can be intuitively understood as
that each pair of the electrodes are closely connected, so that electron spins
flow directly from \textquotedblleft$+$\textquotedblright\ to
\textquotedblleft$-$\textquotedblright\ at one side and are less affected by
the electrode pair at the other side.

A last remarkable point for Fig. \ref{FIG3} is the order of magnitude of
$\left\langle S_{z}\right\rangle $. Strictly speaking, one should refer the
local spin density $\left\langle S_{z}\right\rangle $ to the spin accumulation
only for those patterns in closed samples. When attaching unbiased transverse
leads, electrons are free to leak out, such that the local spin density shows
merely the spatial spin distribution. It is only when the leads at the
transverse sides are unplugged that those spins deflected by the SO force
accumulate at the transverse edges and that the local spin density showing
obvious peaks near the edges deserve the name spin-Hall accumulation. In
addition, such a leakage of the spins due to plugged leads (not necessary
unbiased) may lower the order of magnitude of $\left\langle S_{z}\right\rangle
$ as can be clearly seen by comparing the three groups in Figs. \ref{FIG3}:
(a)--(c), (d)--(f), and (g)--(h).

\subsection{Competition forces in-between Rashba and Dresselhaus
interactions\label{SymmetryBreaking:RDcoexist}}

The semiclassical force provides a simple description of the SHE. Using the
Heisenberg equation of motion with Hamiltonian $\mathcal{H}$ given by Eqs.
(\ref{Hamiltonian})--(\ref{HD}), the velocity can be obtained as
\begin{align}
\mathbf{v}  &  =\mathbf{\dot{r}=}\frac{1}{i\hbar}\left[  \mathbf{r},H\right]
\nonumber\\
&  =\frac{\mathbf{p}}{m^{\star}}+\frac{1}{\hbar}\left[  \left(  -\alpha
\sigma_{y}+\beta\sigma_{x}\right)  \mathbf{e}_{x}+\left(  \alpha\sigma
_{x}-\beta\sigma_{y}\right)  \mathbf{e}_{y}\right]  , \label{v}%
\end{align}
leading to the SO-coupling-induced force%
\begin{equation}
\mathbf{F}_{\text{SO}}\equiv m^{\star}\mathbf{\dot{v}=}\frac{m^{\star}}%
{i\hbar}\left[  \mathbf{v},H\right]  =\frac{2m^{\star}}{\hbar^{3}}\left(
\alpha^{2}-\beta^{2}\right)  \left(  \mathbf{p\times e}_{z}\right)  \sigma
_{z}. \label{FSO}%
\end{equation}
Apart from the constant prefactor $2m^{\star}/\hbar^{3}$, three transparent
properties of $\mathbf{F}_{\text{SO}}$ given by Eq. (\ref{FSO}) can be
identified. First, spin-up electrons ($\sigma_{z}=+1$) and spin-down electrons
($\sigma_{z}=-1$) feel opposite forces. Second, since $\mathbf{F}_{\text{SO}}$
is always perpendicular to the electron transport direction [$\mathbf{F}%
_{\text{SO}}\propto\left(  \mathbf{p\times e}_{z}\right)  \perp\mathbf{p}$],
no work is done by this force. In other words, the spin current caused by
$\mathbf{F}_{\text{SO}}$ is dissipationless. Third, the Rashba and Dresselhaus
interactions contribute oppositely to $\mathbf{F}_{\text{SO}}$, such that they
compete against each other. Thus the sign of $\mathbf{F}_{\text{SO}}$ is also
governed by $\left(  \alpha^{2}-\beta^{2}\right)  $, implying a reverse of the
entire accumulation pattern when interchanging $\alpha$ with $\beta$.

\begin{figure}
[ptb]
\begin{center}
\includegraphics[
height=3.7507in,
width=3.1938in
]%
{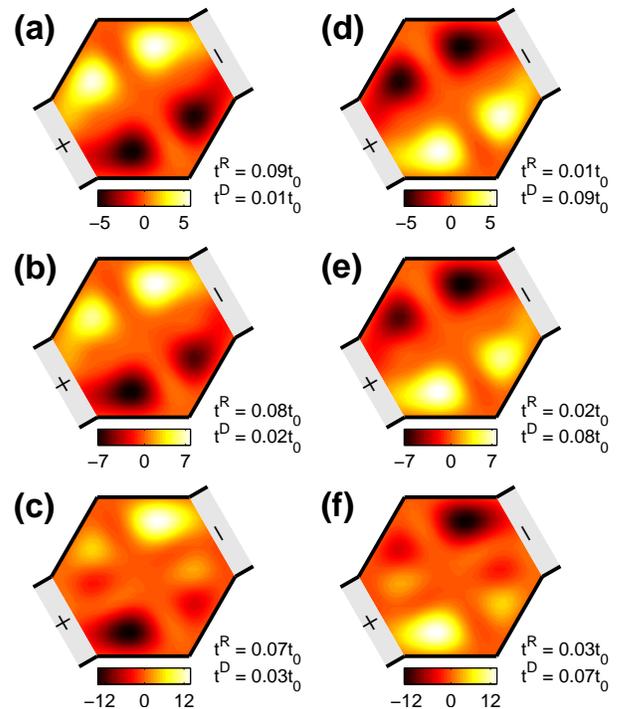}%
\caption{(Color online) Spin-Hall accumulation $\left\langle S_{z}%
\right\rangle $ in units of $\left(  \hbar/2\right)  \times10^{-5}$ for $N=12$
hexagonal samples, each with two head-to-tail leads under low bias, for
various $t^{R}$'s and $t^{D}$'s. Left column (a)--(c) and right column
(d)--(f) are related by interchanging $t^{R}$ and $t^{D}$.}%
\label{FIG4}%
\end{center}
\end{figure}

To demonstrate this SHARE due to the exchange of $\left(  \alpha,\beta\right)
\leftrightarrow\left(  \beta,\alpha\right)  $, or $\left(  t^{R},t^{D}\right)
\leftrightarrow\left(  t^{D},t^{R}\right)  $, let us consider again the
low-biased $N=12$ hexagonal samples with two head-to-tail leads. We show SHA
patterns for various $\left(  t^{R},t^{D}\right)  $ in Fig. \ref{FIG4}, where
the left and right columns are related by interchanging $t^{R}$ and $t^{D}$,
and the SHARE induced by the DSO coupling is clearly seen: patterns with
$\left(  t^{R},t^{D}\right)  \leftrightarrow\left(  t^{D},t^{R}\right)  $ are
reversal of each other. One can also observe that whereas we put $t^{R}>t^{D}$
in the left column and $t^{R}<t^{D}$ in the right column, the reversed
patterns mean the exchange between the spin-up and spin-down states, implying
a sign change in the spin-Hall conductivity as previously
predicted.\cite{Sinova2}%

Another interesting point is that the coexistence of the RSO and DSO couplings
breaks the spin-Hall symmetry. When one of $t^{R}$ or $t^{D}$ dominates, such
as Figs. \ref{FIG4}(a) or \ref{FIG4}(d), the SHA on one side is (nearly) an
mirror (with different sign) of that on the other side; while the magnitude of
$t^{R}$ and $t^{D}$ becomes competitive, such as Figs. \ref{FIG4}(c) or
\ref{FIG4}(f), the spin-Hall symmetry is broken, or distorted. Interestingly,
this distortion effect does not simply kill the SHA pattern. Instead, the spin
accumulation on only one side becomes vague while that on the other side
becomes even stronger, as shown in Figs. \ref{FIG4}(c) or \ref{FIG4}(f).
However, for systems with $t^{R}\approx t^{D}$ vanishing $\left\langle
S_{z}\right\rangle $ inside the 2DEG is obtained (not shown here), as already
implied in Eq. (\ref{FSO}).

\section{Concluding remarks\label{SecConclusion}}

In conclusion, the intrinsic SHE in hexagon-shaped 2DEG is investigated. The
tight-binding model in triangular lattice which suits the hexagon edges is
constructed (within the long wave length limit) for the linear Rashba and
Dresselhaus [001] models. Applying the derived Hamiltonian (with matrix
elements summarized in Table \ref{Hij}) to the LK formalism, we have shown the
SHA patterns in hexagonal samples under various conditions, including SO
interaction strength, bias strength, sample size, lead configuration, and SO
interaction types. The spin-Hall symmetry is identified and discussed in
detail, while the reversal effect of the SHA pattern, which may be induced by
the sample size and the competition between the RSO and DSO couplings, are
discussed. Contrary to the SHA in rectangular samples, we demonstrate here
that the accumulation pattern can be very different due to the geometry effect.

Before closing, three remarks are worthy of mention here. First we highlight
again the main difference between the hexagonal and the rectangular samples.
Although bending the spin-Hall symmetry by applying non-straight biasing can
be done in the rectangular case, transport angles with multiples of 90 degrees
lowered to 60 enlarge the option of lead configuration. Explicitly, the total
number of ways of biasing the sample, e.g., for the two-lead case, is
increased from $C_{2}^{4}=6$ to $C_{2}^{6}=15$, making more detailed analysis
possible. (Note that in general, samples should be treated as
crystallography-dependent, so that each biasing direction is nontrivial.) In
addition, the oblique side boundary varying with longitudinal position making
the SHARE observable [Fig. \ref{FIG3}(c)]. Such a reversal due to spin
precession cannot be observed in one single rectangular sample.

Secondly, we remark on the sample size dependence. There are two
characteristic lengths relavant in our hexagonal samples: spin precession
length $L_{\text{SO}}$ and Fermi wave length $\lambda_{F}$. As we have
discussed the spin-orbit force, the former is the sign criterion of the spin
accumulation pattern. As long as the device size is larger than $L_{\text{SO}%
}$, the direction of the spin-orbit force, and hence the sign of $\left\langle
S_{z}\right\rangle $ accumulating at the two sides, oscillate with the
increase of the sample size. The latter, $\lambda_{F}$, modulates the spatial
distribution of $\left\langle S_{z}\right\rangle $ and thus induces additional
peaks similar to the standing waves. This means that in samples with the
dimension of the order of or larger than $\lambda_{F},$ such waves can be
observed. In our calculations, we have $E_{F}-E_{b}=0.2t_{0}\approx17$ m$%
\operatorname{eV}%
$ and thus $\lambda_{F}=2\pi\hbar/\sqrt{2m^{\star}E_{F}}\approx13a$, which is
slightly shorter than $L_{\text{SO}}$. A good example is to compare Fig.
\ref{FIG2}(b) and Fig. \ref{FIG3}(f). Under exactly the same conditions (lead
configuration, bias strength, etc.), the former with $N=8$ shows no additional
peaks, while in the latter with $N=12$ the $\left\langle S_{z}\right\rangle $
distribution exhibits two peaks at each side due to wave function modulation.

Finally, we stress again the nature of our calculation. It is the hexagonal
boundary that makes discretizing the two-dimensional space inside the sample
into a triangular lattice necessary. In the long wave length limit, where the
Fermi energy is close to the band bottom so that crystal structure information
is less important, our calculation simply reveals the free electron gas
behavior. It is only when the Fermi level no longer approaches the band bottom
that the band structure effect enters and the realistic crystal structure
needs be taken.

Our present work provides a fundamental description of the geometry effect on
the intrinsic SHE, taking the hexagon as the specific case. Additional degrees
of freedom on configuring the attached leads may imply alternative ways of
manipulating spins and even other spin-Hall experiment setups.

\begin{acknowledgments}
This work is supported by the Republic of China National Science Council Grant
No. 95-2112-M-002-044-MY3.
\end{acknowledgments}


\begin{thebibliography}{99}                                                                                               %


\bibitem {App1}S. A. Wolf, D. D. Awschalom, R. A. Buhrman, J. M. Daughton, S.
von Moln\'{a}r, M. L. Roukes, A. Y. Chtchelkanova, and D. M.Treger, Science
\textbf{294}, 1488 (2001).

\bibitem {App2}G. A. Prinz, Science \textbf{282}, 1660 (1998).

\bibitem {Dyakonov}M. I. D'yakonov and V.I. Perel', Zh. Eksp. Teor. Fiz.
\textbf{13}, 657 (1971) [Sov. Phys. JEPT \textbf{33}, 467 (1971)].

\bibitem {Hirch}J. E. Hirsch, Phys. Rev. Lett. \textbf{83}, 1834 (1999).

\bibitem {Zhang}S. Zhang, Phys. Rev. Lett. \textbf{85}, 393 (2000).

\bibitem {Extrinc}Wang-Kong Tse, J. Fabian, I. \v{Z}uti\'{c}, and S. Das
Sarma, Phys. Rev. B. \textbf{72}, 241303(R) (2005)

\bibitem {Y.K.Kato}Y. K. Kato, R. C. Myers,A. C. Gossard, D. D. Awschalom,
Science \textbf{306}, 1117 (2004); J. Wunderlich, B. Kaestner, J. Sinova, and
T. Jungwirth, Phys. Rev. Lett. \textbf{94}, 047204 (2005); V. Sih, R.C. Myers,
Y. K. Kato, W.H. Lau, A. C. Gossard, D. D. Awschalom, Nature Phys. \textbf{1},
31(2005); B. Kaestner, J. Wunderlich, T. Jungwirth, J. Sinova, K. Nomura, A.
H. MacDonald, Phys. E. \textbf{34}, 47(2006).

\bibitem {Th1}S. Murakami, N. Nagaosa, and S. C. Zhang, Science \textbf{301},
1348 (2003)

\bibitem {Th2}S. Murakami, N. Nagaosa, and S. C. Zhang, Phys. Rev. B
\textbf{69}, 235206 (2004).

\bibitem {Th3}S. Murakami, Phys. Rev. B \textbf{69}, 241202(R) (2004).

\bibitem {Th4}D. Culcer, J. Sinova, N. A. Sinitsyn, T. Jungwirth, A. H.
MacDonald, and Q. Niu, Phys. Rev. Lett. \textbf{93}, 046602 (2004).

\bibitem {Th5}L. Hu, J. Gao, and S.-Q. Shen, Phys. Rev. B \textbf{70}, 235323 (2004).

\bibitem {Th6}G. Y. Guo, Yugui Yao, and Q. Niu, Phys. Rev. Lett. \textbf{94},
226601 (2005).

\bibitem {Th7}J. Schliemann and D. Loss , Phys. Rev. B\textbf{ 69}, 165315 (2004).

\bibitem {Th8}A. A. Burkov, Alvaro S. Nunez, and A. H. MacDonald, Phys. Rev. B
\textbf{70}, 155308 (2004).

\bibitem {Th9}S.-Q. Shen, M. Ma, X. C. Xie, and F. C. Zhang, Phys. Rev. Lett.
\textbf{92}, 256603 (2004).

\bibitem {Th10}S.-Q. Shen, Phys. Rev. B. \textbf{70}, 081311(R) (2004).

\bibitem {Th11}Ming-Che Chang, Phys. Rev. B\textbf{ 71}, 085315 (2005).

\bibitem {Sinova1}J. Sinova, D. Culcer, Q. Niu, N.A. Sinitsyn, T. Jungwirth,
and A.H. MacDonald, Phys. Rev. Lett. \textbf{92}, 126603 (2004).

\bibitem {Sinova2}N. A. Sinitsyn, E. M. Hankiewicz, W. Teizer, and J. Sinova,
Phys. Rev. B \textbf{70}, 081312(R) (2004).

\bibitem {Finite1}L. Sheng, D. N. Sheng, and C. S. Ting, Phys. Rev. Lett.
\textbf{94}, 016602 (2005).

\bibitem {Nikolic}B. K. Nikoli\'{c}, S. Souma, L. P. Z\^{a}rbo, and J. Sinova,
Phys. Rev. Lett. \textbf{95}, 046601 (2005).

\bibitem {Nikolic2}B. K. Nikoli\'{c}, L. P. Z\^{a}rbo, and S. Souma, Phys.
Rev. B \textbf{73}, 075303 (2006).

\bibitem {SOforce}B. K. Nikoli\'{c}, L. P. Z\^{a}rbo, and S. Welack, Phys.
Rev. B \textbf{72}, 075335 (2005).

\bibitem {Finite2}B. K. Nikoli\'{c}, L. P. Z\^{a}rbo, and S. Souma, Phys. Rev.
B \textbf{72}, 075361 (2005).

\bibitem {Finite3}K. Nomura, J. Wunderlich, J. Sinova, B. Kaestner, A. H.
MacDonald, and T. Jungwirth, Phys. Rev. B \textbf{72}, 245330 (2005).

\bibitem {QW1}J. Wang and K. S. Chan, Phys. Rev. B \textbf{72}, 045331 (2005).

\bibitem {QW2}K. Hattori and H. Okamoto, Phys. Rev. B \textbf{74}, 155321 (2006).

\bibitem {Rashba}E.I. Rashba, Sov. Phys. Solid State \textbf{2}, 1109 (1960).

\bibitem {Nitta}J. Nitta, T. Akazaki, H. Takayanagi, and T. Enoki, Phys. Rev.
Lett. \textbf{78}, 1335 (1997).

\bibitem {Das2}B. Das, D.C. Miller, S. Datta, R. Reifenberger, W.P. Hong, P.K.
Bhattacharya, J. Singh, and M. Jaffe, Phys. Rev. B \textbf{39}, R1411 (1989).

\bibitem {Dresselhaus}G. Dresselhaus, Phys. Rev. \textbf{100}, 580 (1955).

\bibitem {Lommer}G. Lommer, F. Malcher, and U. R\"{o}ssler, Phys. Rev. B
\textbf{32}, R6965 (1985); Yu. A. Bychkov and E. I. Rashba, in
\textit{Proceedings of the 17th International Conference on Physics of
Semiconductors, San Francisco, 1984} (Springer, New York, 1985), p. 321; M. I.
D'yakonov and V. Y. Kachorovskii, Sov. Phys. Semicond. \textbf{20}, 110 (1986).

\bibitem {Dyakonov2}M. I. D'yakonov and V. I. Perel', Sov. Phys. JETP
\textbf{33}, 1053 (1971).

\bibitem {Bastard}G. Bastard and R. Ferreira, Surf. Sci. \textbf{267}, 335 (1992).

\bibitem {Dattabook}Supriyo Datta, \textit{Electronic transport in mesoscopic
systems }(Cambridge University Press, Cambridge, 1995).

\bibitem {Keldysh}L. V. Keldysh, Sov. Phys. JETP \textbf{20}, 1018 (1965).

\bibitem {Caroli}C. Caroli, R. Combescot, P. Nozieres, and D. Saint-James, J.
Phys. C \textbf{4}, 916 (1971).

\bibitem {KeldyshDerivation}P. Danielewicz, Ann. Phys. (N.Y.) \textbf{152},
239 (1984).

\bibitem {IntroToKeldysh}Robert van Leeuwen, Nils Erik Dahlen, Gianluca
Stefanucci, Carl-Olof Almbladh, and Ulf von Barth, e-print arXiv:cond-mat/0506130.

\bibitem {HaoPL}Ming-Hao Liu and Ching-Ray Chang, Phys. Rev. B \textbf{74},
195314 (2006).
\end{thebibliography}
\end{document}